\documentclass[aps,prd,twocolumn,groupedaddress]{revtex4-1}
\newcommand{\Ref}[1]{(\ref{#1})}
\begin{document}
\title{Lower bound on the magnetic field strength in the hot universe}
\author{E.~Elizalde}
\email[]{elizalde@ieec.uab.es}
\affiliation{Institute for Space Science, ICE-CSIC and IEEC,
Campus UAB, 08193 Bellaterra, Barcelona, Spain}
\author{V.~Skalozub}
\email[]{skalozubv@daad-alumni.de}
\affiliation{Dnipropetrovsk National University, 49010
Dnipropetrovsk, Ukraine}

\date{\today}

\begin{abstract}
It is assumed that long range coherent magnetic fields in the
universe were spontaneously generated at high temperature due to
vacuum polarization of non-Abelian gauge fields, and resulted in
the present intergalaxy magnetic field. The zero value of the
screening mass  for fields of this type was discovered recently.
Here, a procedure to estimate the field strengths at different
temperatures is developed and the lower bound  on the magnetic
field strength $B \sim 10^{14} G$, at    the electroweak phase
transition temperature,   is  derived. As a particular case, the
standard model is considered. Some model dependent peculiarities
of the phenomena under investigation are briefly discussed.
\end{abstract}

\maketitle

\section{Introduction}
Intergalactic magnetic fields are among the most interesting
discoveries in modern cosmology.  Recently, lower bounds of the
order of $B \sim 10^{-15}G$ have been established observationally
\cite{Ando,Neronov} and the search for the origin of these fields
has intensified. One of the best candidates are clearly primordial
fluctuations. But there is also a number of other candidates (for
a review, see \cite{Kunze}).

In this paper we would like to discuss some different mechanism,
based on non-Abelian  magnetic fields. As it was shown recently, a
spontaneous magnetization appears in non-Abelian gauge theories at
high temperature. This was found by analytic methods in
\cite{Starinets:1994vi}--\cite{Skalozub:1999bf} and it was
confirmed by lattice simulations in \cite{Demchik:2008zz}. The
basic idea rests on the known observation that in non-Abelian
gauge theories at high temperature a spontaneous vacuum
magnetization occurs. This is the consequence of the  spectrum of
a color charged gluon,
\begin{equation} \label{spectrum} p^2_{0} = p^2_{||} + (2 n + 1) g
B\qquad(n = - 1, 0, 1,... ), \end{equation}
in a homogeneous magnetic background, $B$, $p_{||}$ is a momentum
directed along the field.  Here, a tachyonic mode is present in
the ground state ($n=-1$).  In fact, one observes that $p_0^2<0$
resulting from the interaction of the magnetic moment of the
spin-1 field with the magnetic field. This phenomenon was first
observed by Savvidy \cite{Savvidy:1977as} at zero temperature,
$T=0$, and got known as the Savvidy vacuum. However, at zero and
low temperature this state is not stable. It decays under emission
of gluons until the magnetic field $B$ disappears. The picture
changes with increasing temperature where a stabilization sets in.
This stabilization is due to vacuum polarization and it depends on
two dynamical parameters. These are a magnetic  mass of the color
charged gluon and an $A_0$-condensate, which is proportional to
the Polyakov loop \cite{Ebert:1996tj}. This configuration is
perfectly stable, since its energy is below the perturbative one
and the minimum is reached for a field of order $g B \sim g^4
T^2$.  Although the phenomenon was discovered in $SU(2)$
gluodynamics, it is common to other $SU(N)$ gauge fields which can
be used to extend the standard $(SU(2) \times U(1))_{ew} \times
SU(3)_c$ model of elementary particles.

An important property of such temperature dependent magnetic
fields is the vanishing of their magnetic mass,   $m_{magn} = 0$.
This was found   both in one-loop  analytic calculations
\cite{Bordag:2006pr} and also in lattice simulations
\cite{Antropov:2010}. The mass parameter describes the inverse
spatial scales of the transverse field components, similarly to
the Debye mass $m_D$, related to the inverse space scale for the
electric (Coulomb) component. The absence of a screening mass
means that the spontaneously generated Abelian  chromomagnetic
fields are long range at high temperature, as is common for the
$U(1)$ magnetic field. Hence, it is reasonable to believe that
at each stage of the  evolution of the hot universe, spontaneously
created, strong,  long-range magnetic fields of different types
have been present. Owing to the property of being unscreened, they
have influenced various processes and phase transitions.

The  dependence  on the temperature of these fields differs from
that of the typical  $U(1)$ magnetic fields. Recall that, in the
latter case, the magnetic (in fact, hypermagnetic) field, created
by some specific mechanism, is implemented in a hot plasma and
decreases according to the law $B \sim T^2$, which is a
consequence of magnetic flux conservation (see, for instance,
\cite{Kunze}). However, in the case of spontaneous vacuum
magnetization the magnetic flux is {\it not} conserved. Instead, a
specific  flux value is generated at each temperature. This fact
has to be taken into consideration when the cooling pattern of the
hot non-Abelian plasma is investigated. This also concerns the
$SU(2)_{ew}$ component of the electromagnetic field.

In the present paper we estimate the strength of the magnetic
field at the temperatures of the electroweak $T_c^{ew}$ phase
transition, assuming that this field was spontaneously generated
by a mechanism as described above. Although such phenomenon is
nonperturbative, we carry out an actual calculation in the
framework of a consistent effective potential (EP) accounting for
the one-loop, $V^{(1)}$, and the daisy (or ring), $V^{ring}$,
diagram contributions of the standard model.  In Sect.~2 we
qualitatively describe, in more detail, some important aspects of
the investigated phenomena. In Sect.~3 the EP  of an Abelian
constant electromagnetic $B$ field at finite temperature is
obtained. It is then used, in Sect.~4, to estimate the magnetic
field strength at the electroweak  phase transition temperatures.
A discussion of the results obtained in the paper together with
some prospects for further work are provided in Sect.~5.
\section{Qualitative consideration}
In this section we describe, in a qualitative manner, the most relevant
aspects of the phenomena considered.  All of them follow from very basic
asymptotic freedom and spontaneous symmetry breaking considerations at
finite temperature.  Our main assumption is that the intergalactic
magnetic field has been spontaneously created at high temperature.
We believe this to be a quite reasonable idea because, physically, the
magnetization is the consequence of a large magnetic moment for
charged non-Abelian gauge fields (let us just remind of the gyromagnetic
ratio $\gamma = 2$ for $W$-bosons). This property eventually results in the
asymptotic freedom of the model in the presence of external fields.
We will discuss the procedure to relate the present value of the
intergalactic magnetic field with the one generated in the restored phase.

First, we note that in non-Abelian gauge theories magnetic flux
conservation does not hold. This is due to spontaneous  vacuum
magnetization which depends on the temperature. The vacuum  acts
as a specific source  generating classical fields. Second, the
magnetization is strongly dependent on the scalar field condensate
present in the vacuum  at low temperature. This point was
investigated at zero temperature by Goroku \cite{Goroku}.  For
finite temperature, it is considered in the present paper for the
first time.  The observation is that, in both cases, the
spontaneous vacuum magnetization takes place for a small scalar
field $\phi \not = 0$, only. For the values of $\phi$
corresponding to any first order phase transition  it  does not
happen. This means that, after the electroweak phase transition
occurs, the vacuum polarization  ceases to generate   magnetic
fields and magnetic flux conservation holds again.  As a result,
the familiar dependence on the temperature $B \sim T^2$  is
restored.

Another aspect of the problem is the composite structure of the
electromagnetic field $A_{\mu}$.  The potentials read
\begin{eqnarray} \label{AZ}
A_\mu &=& \frac{1}{\sqrt{g^2 + g'^2}} ( g' A^3_\mu + g b_\mu ), \nonumber\\
Z_\mu &=& \frac{1}{\sqrt{g^2 + g'^2}} ( g A^3_\mu - g' b_\mu ),
\end{eqnarray}
where $Z_\mu $ is the $Z$-boson potential, $A^3_\mu, b_\mu$ are
the Yang-Mills gauge field  third projection in the weak isospin
space and the potential of the hypercharge gauge fields,
respectively, and $g$ and $g'$ are correspondingly $SU(2)$ and
$U(1)_Y$ couplings. After the electroweak phase transition, the
$Z$-boson acquires mass and the field is screened. Since the
hypermagnetic field is  not spontaneously generated, at high
temperature only the component $A_\mu = \frac{1}{\sqrt{g^2 +
g'^2}}  g' A^3_\mu = \sin \theta_w  A^3_\mu$ is present. Here
$\theta_w $ is the Weinberg angle, $\tan \theta_w = \frac{g'}{g}$.
This is the only component responsible for the intergalactic
magnetic field at low temperature.

In the restored phase the field $b_\mu = 0$, and the complete
weak-isospin chromomagnetic field $A^{(3)}_\mu$ is unscreened.
This is because  the magnetic mass of this field is zero
\cite{Antropov:2010}. Thus, the field is a long range one. It
provides  the coherence length to be sufficiently  large. After
the phase  transition, part of the field is screened by the scalar
condensate. In the restored phase, the constituent of the weak
isospin field corresponding to the magnetic one  is  given by the
expression
\begin{equation} \label{fieldT} B(T) = \sin \theta_w (T) B^{(3)}(T),\end{equation}
where $B^{(3)}(T)$ is the  strength of the field generated spontaneously.

To relate the present value of the intergalactic magnetic field with the field which existed before the electroweak phase transition, we take into consideration  that, after the phase transition, the spontaneous vacuum magnetization does not occur.
 Therefore, for the electroweak transition temperature $T_{ew}$ we can write:
\begin{equation} \label{relation} \frac{B(T_{ew})}{B_0} = \frac{T^2_{ew}}{T^2_0} = \frac{\sin \theta_w (T_{ew}) B^{(3)}(T_{ew})}{B_0}. \end{equation}
Here $B_0$ is the present value of the intergalaxy magnetic field
strength $B_0 \sim 10^{- 15} G$. The left-hand-side relates the
value $B(T_{ew})$ with  $B_0$. The right-hand-side allows to
express the weak isospin magnetic field in the restored phase
through $B_0$, knowing the temperature dependence of the Weinberg
angle $\theta_w(T)$. This relation  contains an arbitrary
temperature normalization parameter $\tau$. It can be fixed for
given temperature and $B_0$.  After that, the field strength
values at various temperatures can be calculated. In particular,
the total weak isospin field strength is given by the sum $ \cos
\theta_w (T_{ew}) B^{(3)}(T_{ew}) + B(T_{ew})$.

An important aspect of this scenario is that the precise nature of
the till now unknown theory extending the standard model is not
very important for estimating the field strength $B$ at
temperatures close to $T_{ew}$. This is because any new gauge
field of the extended model in question will be  screened  at the
relevant higher temperatures corresponding to the  spontaneous
symmetry breaking of some basic symmetries.  At high temperatures,
when these symmetries are restored, related magnetic fields
emerge. Using these ideas the value of the field strength at the
Planck era has been estimated by Pollock \cite{Pollock}. Summing
up, we conclude that our estimate here gives a lower bound on the
magnetic field strength for the hot universe.
\section{Effective potential at high temperature}
As we noted above, spontaneous vacuum magnetization and the
absence of a magnetic mass for the Abelian magnetic fields are
nonperturbative effects to be precisely determined, in particular,
in lattice simulations \cite{Demchik:2008zz,Antropov:2010}. The
main conclusions of these investigations are that a stable
magnetized vacuum does exist at high temperature and that the
magnetic mass of the created field is zero. Concerning the actual
value of the field strength, it is close to the one calculated
within the consistent effective potential which takes into account
one-loop plus daisy diagrams. Thus, in the present investigation
we restrict ourselves to that approximation. The main purpose  for
doing this is to be able to develop analytic calculations, in
order to clarify the results obtained.

The complete EP for the standard model is given in the review
\cite{Demchik:1999}. In the present investigation we are
interested in two  different limits:
\begin{enumerate}\item The weak magnetic field and large scalar field condensate, $h = eB/M_w^2 < \phi^2, \phi = \phi_c/\phi_0, \beta = 1/T$. \item The case of restored symmetry, $\phi = 0, g B \not = 0, T \not = 0$. \end{enumerate}For the first case we show the absence of spontaneous vacuum magnetization at finite temperature. For the second one we estimate the field strength at high temperature. Here $M_w $ is the $W$-boson mass at zero temperature, $\phi_c $ is a scalar field condensate, and $\phi_0$  its value at  zero temperature.

To demonstrate the first property we consider the one-loop
contribution of  $W$-bosons (see also Eq.~(27) of
Ref.~\cite{Skalozub:1996ax}),
\begin{eqnarray} \label{L2t}
V^{(1)}_w(T,h,\phi) &=& \frac{h}{\pi^2 \beta^2} \sum\limits_{n = 1}^{\infty} \Bigl[
\frac{(\phi^2 - h)^{1/2}\beta}{n} K_1(n \beta (\phi^2 - h)^{1/2}) \nonumber\\
&-& \frac{(\phi^2 + h)^{1/2}\beta}{n} K_1(n \beta (\phi^2 +
h)^{1/2})\Bigr]. \end{eqnarray}
Here $n$ labels discrete energy values and $K_1(z)$ is the
MacDonald function.

The main goal of our investigation is the restored phase of the
standard model. To this end, we deduce the high temperature
contribution of the complete effective potential relevant for this
case using the results in Ref.~\cite{Demchik:1999}. First, we
write down the one-loop $W$-boson contribution as the sum of the
pure Yang-Mills weak-isospin part $(\tilde{B}\equiv B^{(3)}$),
\begin{eqnarray} \label{VW} V^{(1)}_w(\tilde{B}, T) &=& \frac{\tilde{B}^2}{2}
+ \frac{11}{48} \frac{g^2}{\pi^2} \tilde{B}^2 \log
\frac{T^2}{\tau^2} - \frac{1}{3} \frac{( g \tilde{B})^{3/2} T
}{\pi} \nonumber\\  &-& i \frac{( g \tilde{B})^{3/2} T }{2 \pi} +
O (g^2 \tilde{B}^2), \end{eqnarray}
where $\tau$ is a temperature normalization point, and the charged
scalars  \cite{Skalozub:1996ax},
\begin{equation} \label{Vscalar} V^{(1)}_{sc}(\tilde{B}, T) =  - \frac{1}{96} \frac{g^2}{\pi^2} \tilde{B}^2 \log \frac{T^2}{\tau^2} + \frac{1}{12} \frac{( g \tilde{B})^{3/2} T }{\pi} + O (g^2 \tilde{B}^2), \end{equation}
describing the contribution of longitudinal vector components. The
first term in Eq.~\Ref{VW} is the tree-level energy of the field.
This representation is convenient for the case of extended models
including other gauge and  scalar fields. Depending on the
specific case,  one can take into consideration  the parts
\Ref{VW} or \Ref{Vscalar}, correspondingly. In the standard model,
the contribution of Eq.~\Ref{Vscalar} has to be included with a
factor 2, due to two charged scalar fields entering the scalar
doublet of the model. In the case of the Two-Higgs-Doublet
standard model, this factor must be 4, etc. The imaginary part is
generated because of the unstable mode in the spectrum
\Ref{spectrum}. It is canceled by the term appearing in the
contribution of the daisy diagrams for the unstable mode
\cite{Skalozub:1999bf},
\begin{equation} \label{ringdaisy} V_{unstable} = \frac{g \tilde{B} T}{2 \pi} [\Pi(\tilde{B}, T, n = - 1) - g \tilde{B} ]^{1/2} + i \frac{(g \tilde{B})^{3/2} T}{2 \pi}. \end{equation}
Here $\Pi(\tilde{B}, T, n = - 1)$ is the mean value  for the
charged gluon polarization tensor taken in the ground state $ n =
- 1$ of the spectrum \Ref{spectrum}. If this value is sufficiently
large, spectrum stabilization due to the radiation correction
takes place. This possibility formally follows from the
temperature and field dependences of the polarization tensor in
the high temperature limit $T \to \infty $ \cite{Bordag08}:
$\Pi(\tilde{B}, T, n = - 1)= c~ g^2 T \sqrt{g \tilde{B}} $, where
$c > 0$ is a constant which must be calculated explicitly. At high
temperature the first term can be larger then $g \tilde{B}$.

From Eqs.~\Ref{VW} and \Ref{ringdaisy} it follows that the
imaginary part cancels. Hence, we see that taking rings into
account leads to vacuum stabilization even if $ \Pi(\tilde{B}, T,
n = - 1)$ is smaller then $g \tilde{B}$. Actually, in the latter
case, the imaginary part will be smaller than in Eq.~\Ref{VW}.

The high temperature limit of the fermion contribution looks as
follows,
\begin{equation} \label{fermionEP} V_{fermion} = - \frac{\alpha}{\pi} \sum\limits_{f} \frac{1}{6} q^2_f \tilde{B}^2 \log\frac{ T}{\tau} , \end{equation}
where the sum is extended to all leptons and quarks, and $q_{f}$
is the fermion electric charge in positron units. Hence, it
follows that in the restored phase all the fermions give the same
contribution.

Now, let us present the EP for ring diagrams  describing the long
range correlation corrections at finite temperature
\cite{Carrington:1992,Demchik:2003},
\begin{eqnarray} \label{Vring} V_{ring} &=& \frac{1}{24 \beta^2} \Pi_{00}(0)
- \frac{1}{12 \pi \beta} Tr [\Pi_{00}(0)]^{3/2}\nonumber\\
&+& \frac{(\Pi_{00}(0))^2}{32 \pi^2} \left[\log\bigl(\frac{4
\pi}{\beta (\Pi_{00}(0))^{1/2}}\bigr) + \frac{3}{4} - \gamma
\right],~ \end{eqnarray}
where the trace means summation over all the contributing states,
$\Pi_{00} = \Pi_{\phi}(k = 0, T, B)$ for the Higgs particle;
$m_D^2 = \Pi_{00} = \Pi_{00}(k = 0, T, B)$ are the zero-zero
components of the polarization functions of gauge fields in the
magnetic field taken at zero momenta, called the Debye mass
squared, $\gamma$ is Euler's gamma. These terms are of order $\sim
g^3 (\lambda^{3/2})$ in the coupling constants. The detailed
calculation of these functions is given in
Ref.~\cite{Demchik:1999}. We give the results for completeness,
\begin{eqnarray} \label{Piscalar}
\Pi_{\phi}(0) &=& \frac{1}{24 \beta^2} \left[ 6 \lambda + \frac{6
e^2}{\sin^2 (2\theta_w)} + \frac{3 e^3}{\sin^2 \theta_w} \right]
\nonumber\\ &+& \frac{2 \alpha}{\pi} \sum\limits_f \left(
\frac{\pi^2 K_f}{3 \beta^2} - |q_f B| K_f \right) \nonumber\\ &+&
\frac{(e B)^{1/2}}{8 \pi \sin^2 \theta_w \beta} e^2 3 \sqrt{2}
\zeta\left(-\frac{1}{2}, \frac{1}{2}\right). \end{eqnarray}
Here $K_f = \frac{m_f^2}{m_w^2} = \frac{G^2_{Yukawa}}{g^2}$ and $
\lambda $ is the scalar field coupling. The terms \mbox{$\sim
\!T^2$} give standard contributions  to the temperature mass
squared coming from the boson and fermion sectors. The
$B$-dependent terms are negative (the value of $3 \sqrt{2}
\zeta(-\frac{1}{2}, \frac{1}{2}) = - 0.39$). They  decrease the
value of the screening mass at high temperature. The Debye masses
squared for the photons, $Z$-bosons and neutral current
contributions are, correspondingly,
\begin{eqnarray} \label{mAZ}
m^2_{D, \gamma} &=& g^2 \sin^2\theta_w \left[\frac{1}{3 \beta^2} + O(e B \beta^2)\right], \nonumber\\
m^2_{D, Z} &=& g^2 \left(\tan^2\theta_w + \frac{1}{4 \cos^2\theta}
\right) \left[\frac{1}{3 \beta^2} + O(e B \beta^2)\right],
\nonumber\\ m^2_{D,neutral} &=& \frac{g^2}{ 8 \cos^2\theta_w
\beta^2} (1 + 4 \sin^4 \theta_w) + O(e B \beta^2). \end{eqnarray}
As one can see, the dependence on $B$ appears at order $O(T^{-2})$.

The $W$-boson contributions to the Debye mass of the photons is
\begin{equation} \label{mW} m^2_{D, W} = 3 g^2 \sin^2\theta_w \left[\frac{1}{3 \beta^2} - \frac{(g \sin \theta_w B)^{1/2}}{2 \pi \beta}\right]. \end{equation}
An interesting feature of this expression is the negative sign of
the next-to-leading terms dependent on the field strength.
Finally, we give the contribution of the high temperature part in
Eq.~\Ref{ringdaisy} $\Pi(\tilde{B}, T, n = - 1)$
\cite{Demchik:1999},
\begin{eqnarray} \label{Piunstable}
\Pi(\tilde{B}, T, n = - 1) &=& \alpha \left[12.33 \frac{(g \sin
\theta_w B)^{1/2}}{\beta}
 \right.\nonumber\\ && \left.
 + 4i \frac{(g \sin \theta_w B)^{1/2}}{\beta} \right].
\end{eqnarray}
This expression has been calculated from the one-loop $W$-boson
polarization tensor in the external field at high temperature. It
contains the imaginary part which comes from the unstable mode in
the spectrum \Ref{spectrum}. Its value is small, as compared to
the real one. It  is of the order of the usual damping constants
in plasma at high temperature. Thus, it will be ignored in actual
calculations, in what follows. In fact, this part should be
calculated in a more consistent scheme which starts with a
regularized stable spectrum. On the other hand, as we noted above,
the stability problem is a non-perturbative one. The stabilization
can be realized not only by the radiation corrections but also by
some other mechanisms. For example, due to $A_0$ condensation
\cite{Starinets:1994vi} at high temperature. We observed the
stable vacuum state in the lattice simulations
\cite{Demchik:2008zz}. Therefore, we do believe that this problem
has a positive solution. Summing up, we now have all what is
necessary in order to investigate in depth the problem of
interest.

\section{Magnetic field strength at $T_{ew}$ }
We will now show that spontaneous vacuum magnetization does not
occur at finite temperature and for non-small values of the
scalar field condensate $\phi \not = 0$. To this end we notice
that the magnetization is produced by the gauge field
contribution, given by Eq.~\Ref{L2t}. So, we consider the limit of
$\frac{g B}{T^2} \ll 1$ and $\phi^2 > h$. For this case we use the
asymptotic expansion of $K_1(z)$,
\begin{equation} \label{K1asympt} K_1(z) \sim \sqrt{\frac{\pi}{2 z}} e^{- z}
\left( 1 + \frac{3}{8 z} - \frac{15}{128 z^2} + \ldots \right),
\end{equation}
where $ z = n \beta (\phi^2 \pm h)^{1/2}$. We  now investigate
the limit $\beta \to \infty, \frac{T}{\phi} \ll 1$.  The leading
contribution is then given by the first term of the temperature
sum in Eq.~\Ref{L2t}. We can also substitute $(\phi^2 \pm h)^{1/2}
= \phi ( 1 \pm \frac{ h}{2 \phi^2})$. In this approximation, the
sum of the tree level energy and the contribution \Ref{L2t} reads
\begin{equation} \label{L2tasympt}
{V} =
\frac{h^2}{2} - \frac{h^2}{\pi^{3/2}} \frac{T^{1/2}}{\phi^{1/2}}
\left( 1 - \frac{T}{2 \phi} \right) e^{- {\phi}/{T}}.
\end{equation}
The second term is exponentially small and the stationary equation
$\frac{\partial{V}}{\partial{h}} = 0$  admits the trivial solution
$h = 0$.  This estimate can be  easily verified in numeric
calculation of the total effective potential. Hence, we conclude
that, as at zero temperature \cite{Goroku}, after symmetry
breaking the vacuum spontaneous magnetization does not take place.

To estimate the magnetic field strength in the restored phase at
the electroweak phase transition temperature, the total effective
potential obtained in the previous section must be used and the
parameters entering Eq.~\Ref{relation} need to be calculated. This
can be best done numerically. Specifically, we consider here the
contribution to this potential accounting for the one-loop
$W$-boson terms. The high temperature expansion for the EP coming
from charged vector fields is given in Eq.~\Ref{VW}.  Assuming
stability of the vacuum state, we calculate the value of the
chromomagnetic weak isospin field spontaneously generated at high
temperature from Eqs.~\Ref{VW} and \Ref{Vscalar}:
\begin{equation} \label{fieldT1}
\tilde{B}(T) = \frac{1}{16} \frac{g^3}{\pi^2}
\frac{T^2}{\displaystyle\left(1 + \frac{5}{12} \frac{g^2}{ \pi^2}
\log \frac{T}{\tau}\right)^2 }.
\end{equation}
We relate this expression with the intergalactic magnetic field $B_0$.

Let us introduce the standard parameters and definitions,
$\frac{g^2}{4 \pi} = \alpha_s, \alpha =  \alpha_s \sin \theta_w^2,
\frac{(g')^2}{4 \pi} = \alpha_Y$ and $\tan^2 \theta_w(T) = \frac{
\alpha_Y(T)}{ \alpha_s(T)}$, where $\alpha$ is the fine structure
constant. To find the temperature dependence of the Weinberg
angle, the behavior of the hypercharge coupling $g'$  on the
temperature has to be computed. From Eq.~\Ref{Vscalar} it follows
that this behavior is nontrivial. The logarithmic
temperature-dependent term is negative. But, as is well known, in
asymptotically free models this sign will unavoidably be changed
to a positive value due to the contributions of other fields. This
particular value is model dependent and we will not calculate it
in this paper. Instead, for a rough estimate, we replace it with
the zero temperature number: $\sin^2 \theta_w(T) = \sin^2
\theta_w(0)= 0.23$.

For the given temperature of the electroweak phase transition,
$T_{ew}$, the magnetic field is
\begin{equation} \label{B3T} B(T_{ew}) = B_0 \frac{T^2_{ew}}{T^2_0} =
\sin \theta_w (T_{ew}) \tilde{B}(T_{ew}).\end{equation}
Assuming $T_{ew} = 100 GeV = 10^{11} eV$ and $T_0 = 2.7 K = 2.3267
\cdot 10^{- 4} eV$, we obtain

\begin{equation} \label{Bew} B(T_{ew}) \sim 1.85 ~10^{14} G. \end{equation}
This value can be considered as a lower bound on the magnetic
field strength at the electroweak phase transition temperature.

Hence, for the value of $X =  \log \frac{T_{ew}}{\tau}$, we have the equation
\begin{equation} \label{Xew}
B_0 =  \frac{1}{2}  \frac{\alpha^{3/2}  }{\pi^{1/2} \sin^2
\theta_w } \frac{T^2_0}{\displaystyle\left(1 + \frac{5 \alpha}{3
\pi \sin^2 \theta_w }  X\right)^{2}}. \end{equation}
Since all the values here are known, $\log \tau $ can be estimated.
After that the field strengths at different higher temperatures
can be found. Of course, our estimate is a rough one because of having ignored
the temperature dependence of the Weinberg angle. To guess the
value of the parameter $\tau$ we take  the field strength $B_0
\sim 10^{- 9} G$, usually used in cosmology (see, for example,
\cite{Pollock}). In this case, from Eq.~\Ref{Xew} we obtain $\tau
\sim 300 eV$. For the lower bound value $B \sim 10^{- 15}G$ this
parameter is much smaller. The strong suppression of the field
strength is difficult to explain within the standard model. This
point will be discussed below.

To take into consideration the fermion contribution
Eq.~\Ref{fermionEP}, we have to substitute the expression
$\frac{5}{12} \frac{g^2}{ \pi^2} \log \frac{T}{\tau}$ in
Eqs.~\Ref{fieldT1} and \Ref{Xew} with the value
\begin{equation} \label{plusferm}
\left(\frac{5}{3} -  \sum\limits_{f} \frac{1}{6} q^2_f
\right)\frac{\alpha_s}{ \pi} \log \frac{T}{\tau}.
\end{equation}
In the above estimate, we have taken into account the one-loop
part of the EP of order $\sim g^2$ in the coupling constant. The
ring diagrams are of order $\sim g^3$ and provide a small numeric
correction to  this result in the high temperature approximation.
As it was mentioned before, had we taken into account all the
terms listed in the previous section the results would have not
changed essentially.

The field strength at higher temperatures depends on the
particular model extending the standard one. Spontaneous vacuum
magnetization in the minimal supersymmetric standard model has
been investigated in Ref.~\cite{Demchik:2003}. The field strength
generated in this model is smaller, as compared with the situation
here considered.   Pollock   \cite{Pollock} has investigated this
problem   for the case of the Planck era, where magnetic fields
of  order $B \sim 10^{52} G$ have been estimated. We will further
discuss these results in the concluding section.

\section{Discussion}
Here we summarize our main results. The key issue in the problem
under investigation is the spontaneous magnetization of the
vacuum,  which eliminates the magnetic flux conservation principle
at high temperatures. This vacuum polarization is responsible for
the value of the field strength $B(T)$ at each temperature and
serves as a source for it. We have also shown that, at finite
temperature and after the symmetry breaking, a scalar field
condensate suppresses the magnetization.  Hence it follows that
the actual nature of the particular extended model is not
essential at sufficiently low temperatures when  the decoupling of
heavy gauge fields has happened already. These  statements are new
and come as an interesting surprise, as compared with the standard
notions based on the ubiquitous scenario with magnetic flux
conservation.   In the latter case one assumes that the magnetic
field is created by some mechanism at different stages of the
universe evolution. Then the temperature dependence ($B \sim T^2$)
is regulated by magnetic flux conservation, only.

The present value of the intergalactic magnetic field is related
in our model with  the field strengths at high temperatures in the
restored phase. Because of the zero magnetic mass for Abelian
magnetic fields, as discovered recently \cite{Antropov:2010},
there is no problem in the generation of fields having a large
coherence length.  Knowing the particular properties of the
extended model it is possible to estimate the field strengths at
any temperature. This can be done for different schemes of
spontaneous symmetry breaking (restoration) by taking into account
the fact that, after the decoupling of some massive gauge fields,
the corresponding magnetic fields are screened. Thus, the higher
the temperature is the larger will be the number of strong long
range magnetic fields of different types that will be generated in
the early universe.

Now, let us compare our results with those of Ref.~\cite{Pollock},
where spontaneous vacuum magnetization at high temperature was
applied to estimate the field strength at the Planck era. In that
paper, in order to estimate the field strength, heterotic
superstring theory $E_8 \times E_8^{'}$ was considered as a basic
ingredient. At the Planck era, the magnetic field strength  has
been estimated to be of order $\sim 10^{52} G$. In contrast to our
considerations, it was assumed there that the magnetic field
approximately scales as $B \sim T^2$. That is, vacuum
magnetization was taken into account only at the very first
moments of the universe evolution. Further, recent results
implying a zero magnetic mass for the Abelian chromomagnetic
fields also change the picture of the magnetized early universe
substantially. According to those, the created magnetic fields
existed already on the horizon scales. They were switched off at
some mass scales,  because of  the spontaneous symmetry breaking
as the temperature was lowering  and the decoupling of heavy gauge
fields occured. As a result, at the electroweak phase transition
only the component $B^{(3)}$ of the $SU(2)$ weak isospin group
remains unscreened and eventually results in the present day
intergalactic magnetic field. The processes of decoupling were
also not taken into consideration in Ref.~\cite{Pollock}. Thus, it
was impossible there to relate the electromagnetic field $B_0$
with the magnetic fields generated at high temperatures.

Our analysis has shown that, at the electroweak phase transition
temperature, magnetic fields of the order $B(T_{ew}) \sim 10^{14}
G$ were present. To estimate the field strengths at high
temperatures, one needs to put into play a number of
characteristic features of the standard model and its particular
extension. First,  we note that quarks possess both electric and
color charges. Therefore, there is a mixing between the color and
usual magnetic fields owing to the quark loops. Second, there are
peculiarities related with the  particular content of the extended
model considered. For example, in the Two-Higgs-Doublet standard
model the contribution $\sim (g B)^{3/2} T$ in Eq.~\Ref{VW} is
exactly canceled by the corresponding term in Eq.~\Ref{Vscalar},
because of the four charged scalar fields entering the model. They
interact with gauge fields with the same coupling  constant.
However, in this model the doublets interact differently with
fermions. This changes the effective couplings of the doublets
with gauge fields and results in non-complete cancelations.  As a
result, a strong suppression of the spontaneously created magnetic
field is expected in this model. This, in principle, could explain
a very small value of the  intergalactic magnetic field at low
temperature. There are other peculiarities which influence the
high temperature phase of the universe. They require further
investigations, which  we leave for a future publication.

\begin{acknowledgments}
The authors are indebted with Michael Bordag for numerous
discussions and a reading of the manuscript. One of us (VS) was
supported by the European Science Foundation  activity  "New
Trends and Applicatios of Casimir Effect", and also thanks the
Group of Theoretical Physics and Cosmology, at the Institute for
Space Science, UAB, Barcelona for kind hospitality. EE has been
partly supported by MICINN (Spain), projects FIS2006-02842 and
FIS2010-15640, by the CPAN Consolider Ingenio Project, and by
AGAUR (Generalitat de Ca\-ta\-lu\-nya), contract 2009SGR-994.
\end{acknowledgments}

\end{document}